\documentclass[aps,prl,reprint,amsmath,amssymb,superscriptaddress,floatfix]{revtex4-1}
\usepackage{amsthm,dsfont,ctable,url,braket,mathtools}
\usepackage{natbib}

\usepackage[colorlinks,
            linkcolor=blue,
            anchorcolor=blue,
            citecolor=blue,
urlcolor=blue
            ]{hyperref}

\usepackage{graphicx}
\usepackage[FIGTOPCAP,raggedright,nooneline]{subfigure}
\usepackage{float}
\usepackage{longtable}

\usepackage[T1]{fontenc}
\usepackage{lmodern}

\newcommand{\Tr}{\operatorname{Tr}}

\begin{document}
\title{Experimental self-characterization of quantum measurements}

\author{Aonan \surname{Zhang}}
\altaffiliation{A. Z and J. X contributed equally to this work.}
\affiliation{National Laboratory of Solid State Microstructures, College of Engineering and Applied Sciences and School of Physics, Nanjing University, Nanjing 210093, China}
\affiliation{Collaborative Innovation Center of Advanced Microstructures, Nanjing University, Nanjing 210093, China}

\author{Jie \surname{Xie}}
\altaffiliation{A. Z and J. X contributed equally to this work.}
\affiliation{National Laboratory of Solid State Microstructures, College of Engineering and Applied Sciences and School of Physics, Nanjing University, Nanjing 210093, China}
\affiliation{Collaborative Innovation Center of Advanced Microstructures, Nanjing University, Nanjing 210093, China}

\author{Huichao \surname{Xu}}
\affiliation{National Laboratory of Solid State Microstructures, College of Engineering and Applied Sciences and School of Physics, Nanjing University, Nanjing 210093, China}
\affiliation{Collaborative Innovation Center of Advanced Microstructures, Nanjing University, Nanjing 210093, China}

\author{Kaimin \surname{Zheng}}
\affiliation{National Laboratory of Solid State Microstructures, College of Engineering and Applied Sciences and School of Physics, Nanjing University, Nanjing 210093, China}
\affiliation{Collaborative Innovation Center of Advanced Microstructures, Nanjing University, Nanjing 210093, China}

\author{Han \surname{Zhang}}
\affiliation{National Laboratory of Solid State Microstructures, College of Engineering and Applied Sciences and School of Physics, Nanjing University, Nanjing 210093, China}
\affiliation{Collaborative Innovation Center of Advanced Microstructures, Nanjing University, Nanjing 210093, China}

\author{Yiu-Tung \surname{Poon}}
\affiliation{Department of Mathematics, Iowa State University, Ames, Iowa, IA 50011, USA}
\affiliation{Shenzhen Institute for Quantum Science and Engineering, Southern University of Science and Technology, Shenzhen 518055, China}
\affiliation{Center for Quantum Computing, Peng Cheng Laboratory, Shenzhen, 518055, China}

\author{Vlatko \surname{Vedral}}
\email{phyvv@nus.edu.sg}
\affiliation{Atomic and Laser Physics, Clarendon Laboratory, University of Oxford, Parks Road, Oxford OX13PU, United Kingdom}
\affiliation{Centre for Quantum Technologies, National University of Singapore, 3 Science Drive 2, 117543, Singapore}

\author{Lijian \surname{Zhang}}
\email[]{lijian.zhang@nju.edu.cn}
\affiliation{National Laboratory of Solid State Microstructures, College of Engineering and Applied Sciences and School of Physics, Nanjing University, Nanjing 210093, China}
\affiliation{Collaborative Innovation Center of Advanced Microstructures, Nanjing University, Nanjing 210093, China}

\date{\today}

\begin{abstract}
The accurate and reliable description of measurement devices is a central problem in both observing uniquely non-classical behaviors and realizing quantum technologies from powerful computing to precision metrology. To date quantum tomography is the prevalent tool to characterize quantum detectors. However, such a characterization relies on accurately characterized probe states, rendering reliability of the characterization lost in circular argument. Here we report a self-characterization method of quantum measurements based on reconstructing the response range---the entirety of attainable measurement outcomes, eliminating the reliance on known states. We characterize two representative measurements implemented with photonic setups and obtain fidelities above 99.99\% with the conventional tomographic reconstructions. This initiates range-based techniques in characterizing quantum systems and foreshadows novel device-independent protocols of quantum information applications.
\end{abstract}

\maketitle  
\par
The information of any quantum system we can acquire, manipulate and transmit is finally revealed by quantum measurements. As the measuring devices become increasingly sophisticated, the implementations of both tests of quantum theories and quantum information applications~\cite{Knill2001,RevModPhys.79.135,Higgins2007,PhysRevLett.98.223601} require experimental calibration and certification of measurement apparatus, which is normally achieved by recording the measurement outcomes on probe states. In principle of quantum mechanics, the operation of a quantum measurement on quantum states complies with Born's rule $p_{k}^{(j)}=\Tr (\rho^{(j)} \pi_k),\ k=0,1,...,n-1.$ Here $\{\rho^{(j)}\}$ represent quantum states described by density matrices and $\{\pi_k\}$ is the positive-operator-valued measure (POVM) of a quantum measurement with $n$ outcomes. This formula describes the measurement as a mapping from the state space of quantum systems $\{\rho|\rho \geq 0,\ \Tr(\rho) = 1\}$ to the classically accessible detector outcomes represented in the probability space $\{(p_0,p_1,...,p_{n-1})\}$, thus enabling us to predict the measurement results and also perform the inverse, i.e. to identify the measurement operators in accordance with observed results. To do this, one could probe the measurement device by identical copies of a set of known states, and then find the POVM $\{\pi_k\}$ closest to the observed results, for example, by optimizing the least square function
\begin{equation}
\min\sum_{j,k}\left[p_{k}^{(j)}-\Tr(\rho^{(j)} \pi_k)\right]^2,
\label{eq:qdt}
\end{equation}
under the physical constraint $\pi_k \geq 0$ and $\sum \pi_k=I$, where $I$ denotes the identity operator. This method, known as quantum detector tomography (QDT), has been suggested as the standard tool of characterizing quantum measurements~\cite{PhysRevLett.83.3573,PhysRevLett.93.250407,Lundeen2008,Zhang2012}.
\par
Despite the success of QDT, an unavoidable issue arises in real-world applications, that is, the accuracy of the tomography results relies on precisely calibrated probe states (see Fig.~\ref{fig:scheme:a}). Conversely, to calibrate the source for probe states one requires a convincing measurement device, which forms a fundamental loop paradox. Efforts have been made to develop improved tomography techniques such as self-calibrating tomography, that relaxes partial knowledge in the state or the measurement side~\cite{Bra_czyk_2012,Mogilevtsev_2012}. On the other hand, in certain cases quantum states and measurements can be ``self-tested'' in a device-independent (DI) way~\cite{743501,PhysRevLett.117.260401,PhysRevLett.121.240402,PhysRevA.98.062307}, i.e. without assuming the internal workings of the apparatus used. These self-testing methods originated from ensuring secure cryptography~\cite{743501} and were then utilized to bound dimensionality~\cite{Ahrens2012,Hendrych2012}, generate random numbers~\cite{PhysRevLett.114.150501,Ma2016,PhysRevApplied.7.054018} and verify quantum computers~\cite{Reichardt2013}. In this line, DI tests are typically based on a witness involving observed probabilities so only a specific class of states and measurements can be self-tested within this regime. More recently, there was another idea of DI tests concerning the full attainable range of the input-output correlations~\cite{PhysRevLett.118.250501,DallArno2017a,Agresti_2019}. This provides the possibility of directly inferring the information of the measurement from the range~\cite{DallArno2018} rather than certifying a targeted witness.
\begin{figure}
\centering
    \addtocounter{figure}{1}
  \subfigure{
    \label{fig:scheme:a} 
    \includegraphics[width=\linewidth]{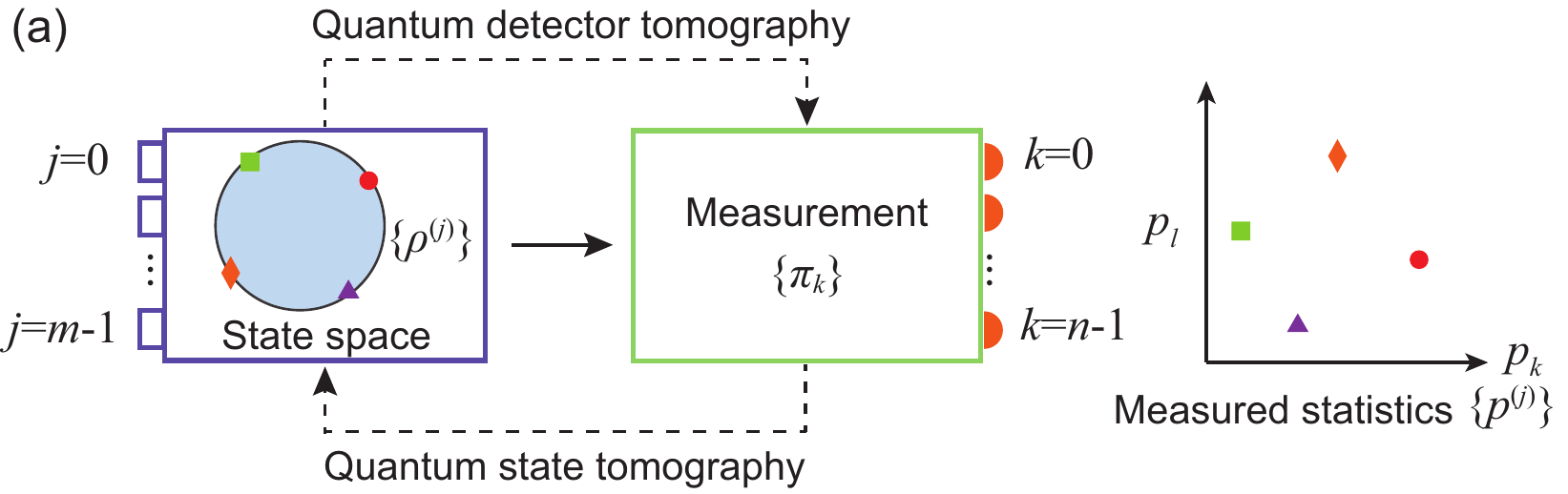}}\\
  \subfigure{
    \label{fig:scheme:b} 
    \includegraphics[width=\linewidth]{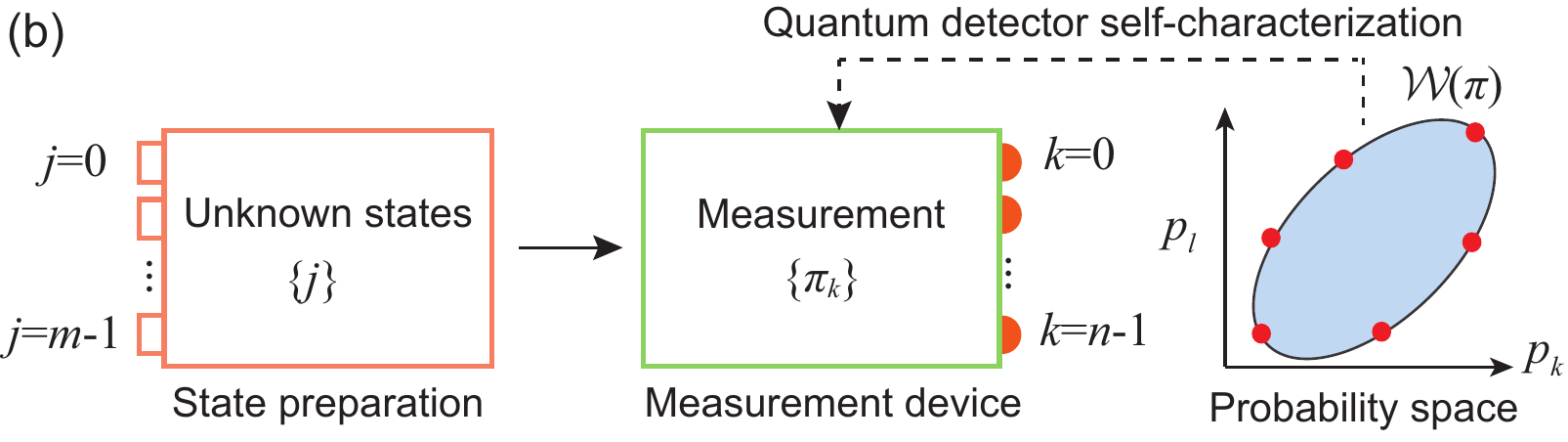}}
  \addtocounter{figure}{-1}
\caption{\label{fig:scheme}Schematic diagram of quantum tomography and self-characterization. (a) Tomography of quantum measurements demands a set of known probe states, whereas tomography of quantum states demands well-calibrated quantum detectors. This forms a loop paradox in calibrating quantum systems. (b) In contrast, quantum detector self-characterization only uses the detector outcome probabilities from the measurement part itself to reconstruct the response range of the measurement, with unknown states, thus can break the aforementioned loop paradox.}
\end{figure}

\par
In this work, we propose and realize quantum detector self-characterization (QDSC), capable of characterizing general unknown quantum measurements, based fully on the detector outcomes of the measurement device itself, thus can break the loop paradox in characterizing quantum systems. The idea is to retrieve the attainable region for measurement results in actual use of the detector, 
$$\mathcal{W}(\pi):=\{(\Tr(\rho\pi_0),...,\Tr(\rho\pi_{n-1}))|\rho\geq 0, \Tr(\rho)=1\},$$
termed as the response range of a quantum measurement. The response range can be formalized as the expectation values of a set of operators and derived by the fundamental constraints on quantum systems and uncertainty relations~\cite{Sehrawat2017}. Distinguished from conventional QDT which explicitly involves probe states, this procedure (conceptually shown in Fig.~\ref{fig:scheme:b}) reconstructs the measurement directly from the statistics of measurement outcomes $\{p^{(j)}\}$, without knowing which states are measured. With practical data in finite statistics, the problem is recast into an optimization problem which aims at giving a best estimation of the range $\mathcal{W}(\pi)$ consistent with the data, that is,
\begin{eqnarray}
\min && \mathcal{F}[\mathcal{W}(\pi),\{p^{(j)}\}], \nonumber \\
\text{subject to~}&& \pi_k\geq 0 \text{~and~} \sum_k \pi_k=I,
\end{eqnarray}
where $\mathcal{F}[\mathcal{W}(\pi),\{p^{(j)}\}]$ is a cost function evaluating how well data fit the estimation. From the estimated range $\mathcal{W}(\pi)$ one can recover the information about the POVM without involving the density matrices of states. Compared with self-calibrating tomography~\cite{Bra_czyk_2012,Mogilevtsev_2012} that combines measurement statistics and priori knowledge in states or measurement operators to perform a joint tomography, the self-characterization method directly analyzes the collective behaviors of the measurement results mapped from the entire state space rather than certain set of states.
\begin{figure}
    \includegraphics[width=\linewidth]{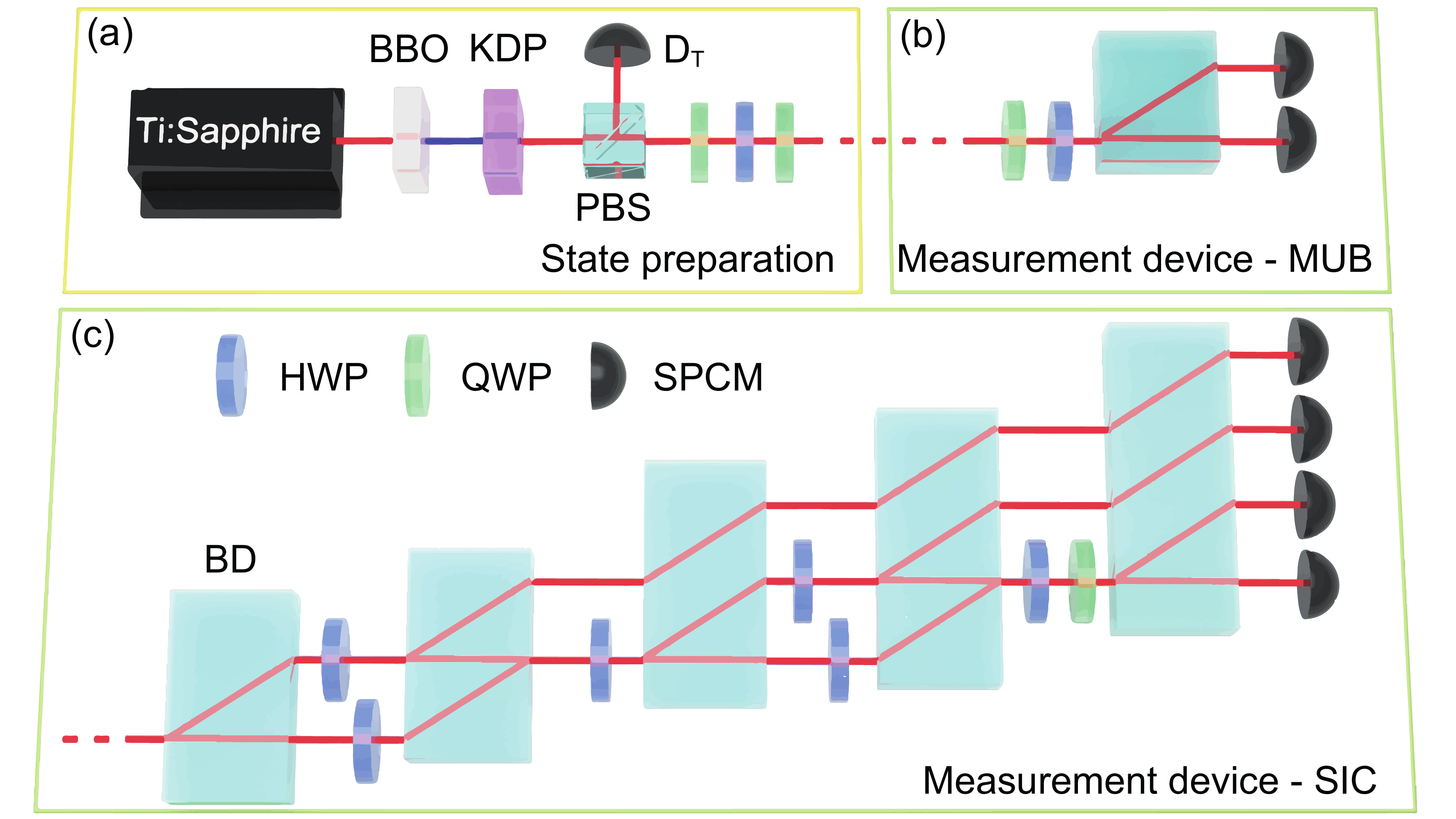}
  \caption{\label{fig:setupsic}Experimental set-up. (a) Heralded single photons are generated via spontaneous parametric downconversion, followed which a set of probe states are prepared by three electronically-controlled waveplates and directed towards the measurement device (b) or (c). (b) The MUB device is composed of two waveplates followed by a beam displacer (BD) to perform projection on a certain basis. (c) The SIC device is a four-outcome general measurement realized by waveplates, BDs and single photon counting modules (SPCMs). BBO, $\beta$-barium borate crystal; KDP, potassium di-hydrogen phosphate; HWP, half wave plate; QWP, quarter wave plate.} 
\end{figure}
\begin{figure*}
    \includegraphics[width=\textwidth]{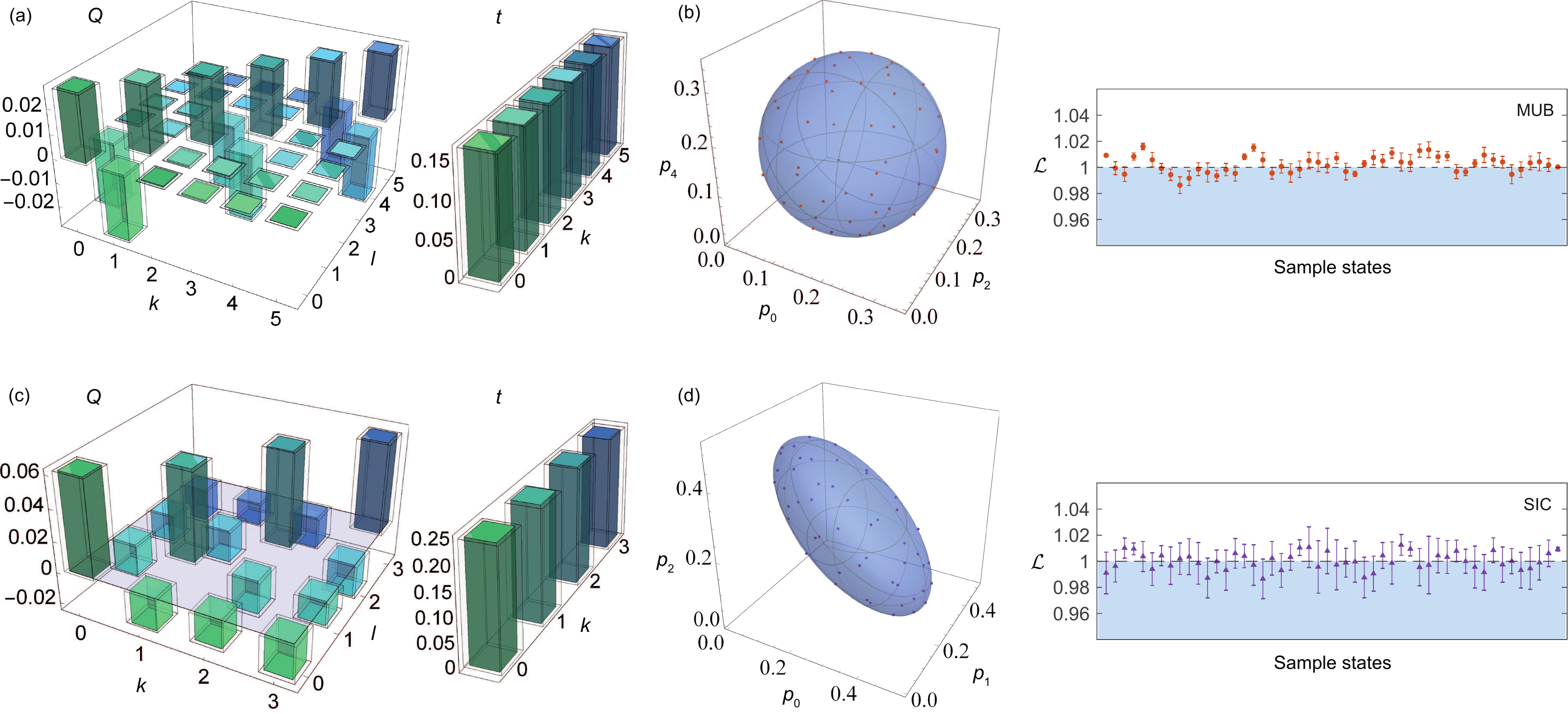}
  \caption{\label{fig:DI}Results of quantum detector self-characterization (QDSC). (a,c) The reconstructed $Q$ and $t$ (chromatic bars) for the MUB and SIC devices respectively. The corresponding results of quantum detector tomography (transparent bars with solid line edges) are also plotted for comparison. (b,d) Left: the estimated response range (blue region) and the measured data (points), illustrated in the probability space despite the linear dependencies of the measurement operators, for the MUB and SIC devices respectively. Right: the detailed results represented in terms of the values of $\mathcal{L}$ in Eq.~(\ref{eq:inequality}). Error bars are standard uncertainties derived from 40 runs of the experiment.} 
\end{figure*}
\par
To apply this QDSC method to the characterization of practical devices, we implemented two representative measurements for tomography purpose, mutually unbiased bases (MUB) and symmetric informationally complete (SIC) measurements for single-qubit system~\cite{Supplementary}, with photonic setups shown in Fig. \ref{fig:setupsic}. These two measurements are of particular interests in quantum information applications~\cite{WOOTTERS1989363,PhysRevA.70.052321}. The experimental set-up consists of two parts: state preparation (a) and measurement (b) or (c). The state preparation starts with a heralded single photon source via spontaneous parametric downconversion. A polarizing beam-splitter and three electronically controlled waveplates prepare probe states $\{\rho^{(j)}\}$ encoded in the polarization degree of freedom of single photons. The states are sent to a measurement apparatus with operations on the polarization modes and spatial modes on the single photons and detection with photon-counting detectors . The clicks of each detector correspond to an outcome $\pi_k$ of the measurement. For both measurements, we collected the measured statistics of detectors for 50 probe states sampled on the Bloch sphere~\cite{Supplementary}. Note although QDSC does not need to know the exact form of probe states, we recorded the settings of state preparation for the following tomographic reconstruction.
\par
For qubit measurements used in our experiment, it has been shown~\cite{PhysRevLett.118.250501} that the response range $\mathcal{W}(\pi)$ is a set $\{p\}$ satisfying
  \begin{equation}
      \mathcal{L}=(p - t)^T Q^+ (p - t) \le 1,
    \label{eq:inequality}
  \end{equation}
and $p$ is subject to $(I - Q Q^+) (p - t) = 0$ which is equivalent to the requirement of linear dependencies among outcomes of the POVM (see Supplemental Material~\cite{Supplementary} for a derivation of Eq.~(\ref{eq:inequality})). The matrix $Q$ and the vector $t$ are given by $Q_{k, l} = \Tr(\pi_{k} \pi_{l})/2- \Tr(\pi_{k}) \Tr(\pi_{l})/4$ and $t_k=\Tr(\pi_k)/2$, and $(.)^+$ denotes the Moore-Penrose pseudoinverse. More precisely, the matrix $Q$ quantifies the overlap of POVM elements and the vector $t$ represents the weight of POVM elements, thus $Q$ and $t$ identify the POVM $\{\pi_k\}$ up to the equivalence class of unitary operations and relabelling of outcomes~\cite{Supplementary}. The physical constraint $\pi_k\geq 0$ can be written as $t_k^2-Q_{k,k}\geq 0$ in the $Q,t$ representation. Geometrically, the inequality is in a center form of an $n$-dimensional (hyper-)ellipsoid centered on $t$. Upon considering the linear dependencies of the POVM elements, Eq. (\ref{eq:inequality}) may reduce to an ellipsoid, an ellipse or a segment depending on the number of linear independent operators in $\{\pi_k\}$.
\par
The characterization in our experiment is based on several assumptions: (i) the dimension of the system (qubit system in our case); (ii) the probe states are adequately sampled to cover the boundary of the state space. In this sense our method is semi-device-independent. In addition, we assume fair sampling, i.e., the registered statistics is a representative sample of the generated states and the state preparation and measurement device are uncorrelated. These requirements are reasonable for an optical experiment and not more than a standard tomography scenario. The characterization procedure firstly extracts features in the data set via singular value decomposition and principle component analysis. This step removes the redundant linear dependent outcomes and is robust against experimental noise (see Supplemental Material for details~\cite{Supplementary}). Then we perform a convex hull of the processed data to obtain the boundary data set $\mathcal{B}$. In the estimation we resort to the direct least squares between the boundary of the estimated range and the boundary data $[1-(p^{(j)}- t)^T  Q^+ (p^{(j)}-t)]^2$ for $j \in \mathcal{B}$ as the cost function. As a result, the characterization is conducted with only the measured statistics by solving the constrained optimization problem
\begin{eqnarray}
\min \sum_{j \in \mathcal{B}} && [1-(p^{(j)}- t)^T  Q^+ (p^{(j)}-t)]^2, \nonumber \\
\text{subject to}
&&~t_k^2-Q_{k,k}\geq 0.
\label{eq:opt}
\end{eqnarray}
\begin{figure}
  \centering
  \addtocounter{figure}{1}
    \subfigure{
    \label{fig:comparison:a} 
    \includegraphics[width=\linewidth]{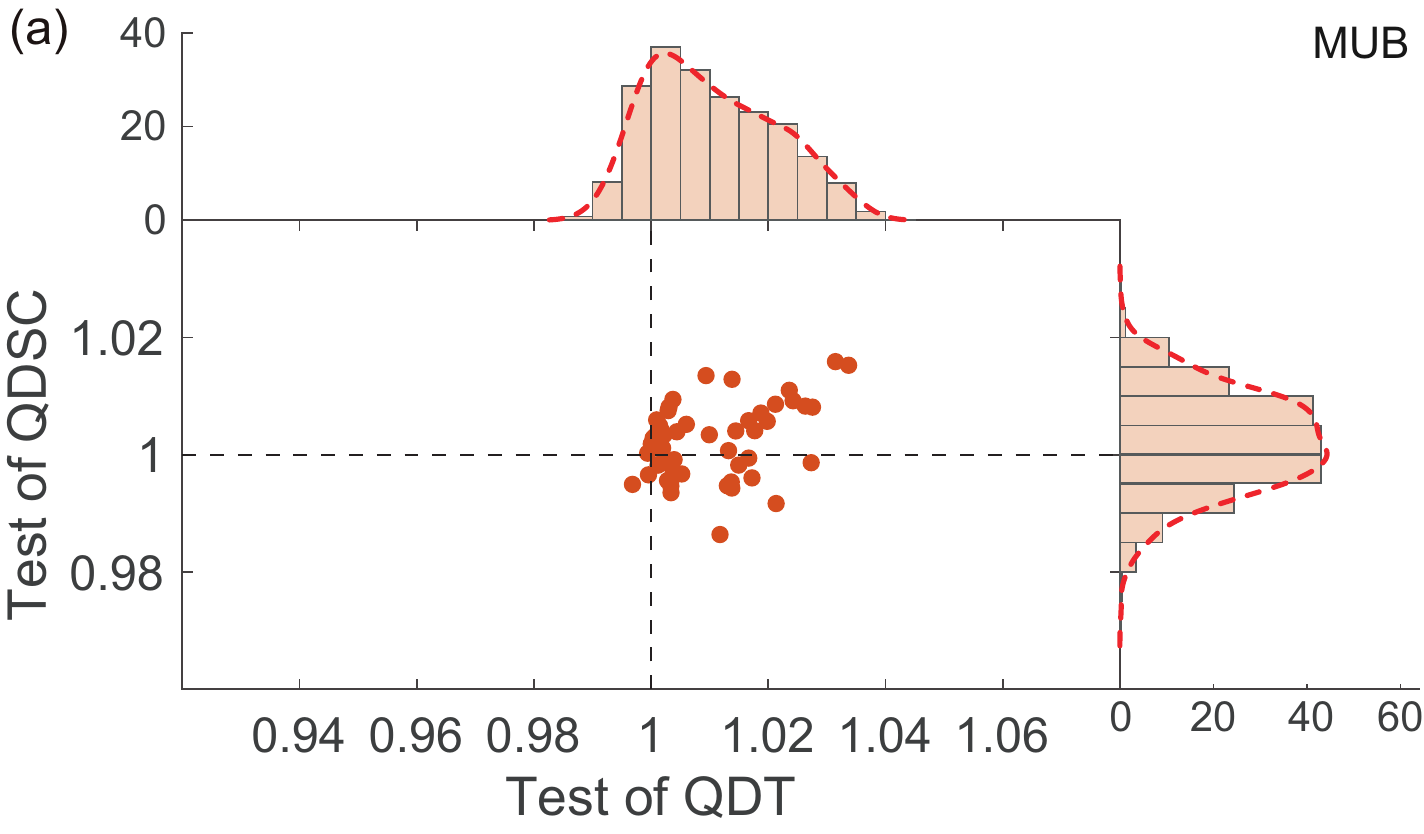}}\\
  \subfigure{
    \label{fig:comparison:b} 
    \includegraphics[width=\linewidth]{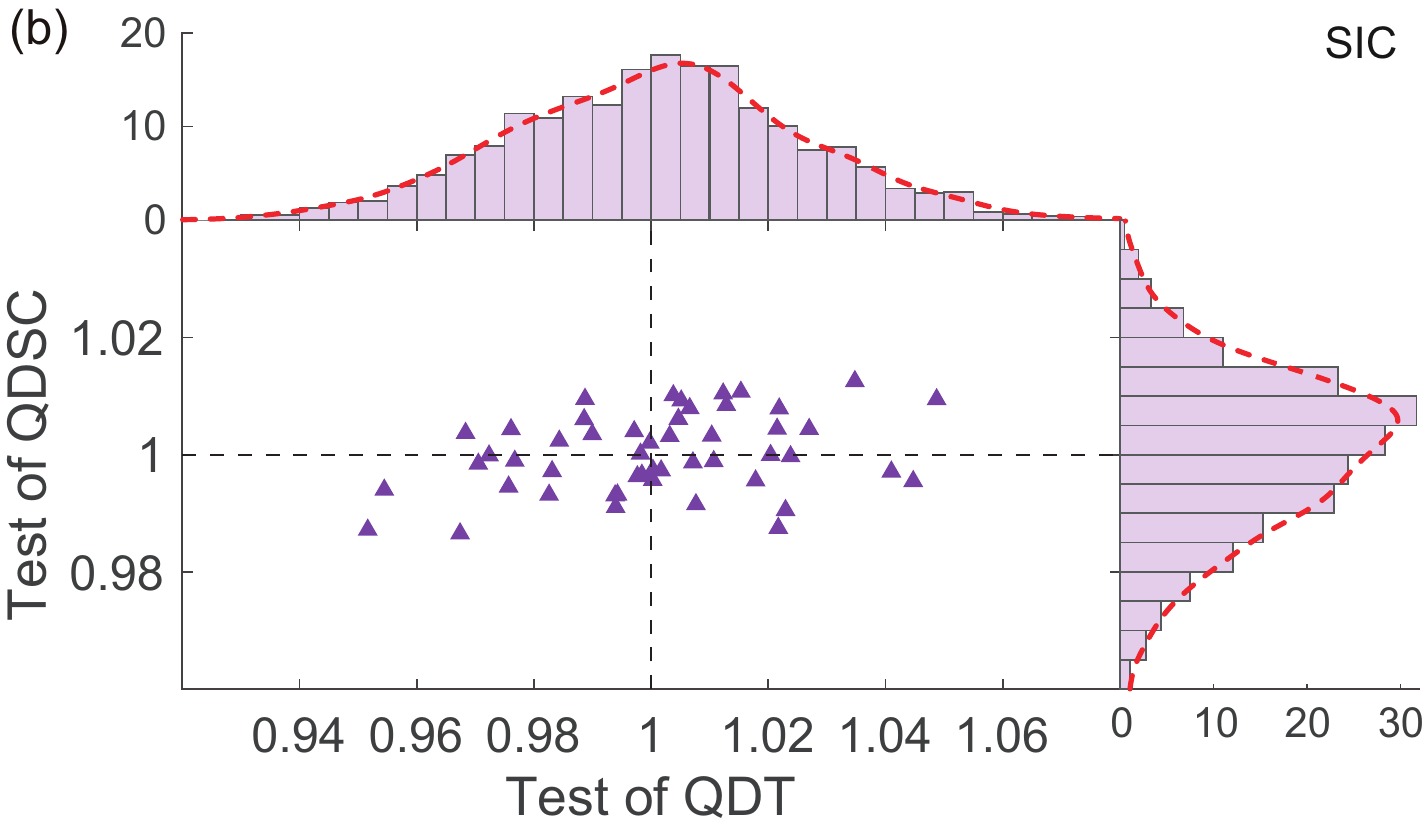}}
  \addtocounter{figure}{-1}
  \caption{Comparison of quantum detector tomography (QDT) and self-characterization (QDSC). The results are represented via tests of (a) the MUB device (red dot) and (b) the SIC device (purple triangle). Each marker represents the measured values of $\mathcal{L}$ in Eq. (\ref{eq:inequality}) averaged over 40 runs for a same probe state. The marginal distributions in the horizontal (QDT) and vertical (QDSC) axes, represented by histograms from the 50$\times$40 data before average and the corresponding Kernel fittings (red dashed lines), reflect the deviations of the measured data from the bound (black dashed lines).}
  \label{fig:comparison} 
\end{figure}
\par
Figure \ref{fig:DI} shows the experimental results of QDSC of the two measurements. To show the performance of self-characterization, we also give the results reconstructed with conventional QDT (with the same probe states) for comparison. In the QDT scenario, we use the measured statistics $\{p^{(j)}\}$, combined with the density matrices $\{\rho^{(j)}\}$ derived by the settings of waveplates, to numerically solve the convex optimization problem in Eq. (\ref{eq:qdt}) and reconstruct the POVM elements $\{\pi_k\}$ (thereby $Q^{\text{tomo}}$ and $t^{\text{tomo}}$). The reconstructed results $Q^{\text{sc}}$ and $t^{\text{sc}}$ via QDSC are in well agreements with the reconstruction by conventional QDT (see Figs. \ref{fig:DI}(a) and \ref{fig:DI}(c)), having the fidelities $F_Q=\left[\Tr\left(\sqrt{Q^{\text{tomo}}}Q^{\text{sc}}\sqrt{Q^{\text{tomo}}}\right)\right]^2/\left[\Tr\left(Q^{\text{tomo}}\right)\Tr\left(Q^{\text{sc}}\right)\right]$ and $F_t=\left(\sum_k \sqrt{t_k^\text{tomo}t_k^\text{sc}}\right)^2$ above 99.99\% with the tomographic reconstruction (QDT) for both implementations~\cite{Supplementary}. To further visualize the results of the QDSC, we plot the reconstructed response range together with the measured data in Figs. \ref{fig:DI}(b) and \ref{fig:DI}(d). The response range and the measured data are illustrated in a three-dimensional probability space of its linear independent outcomes, despite the linear dependent ones (due to the fact that $p_k+p_{k+1}=1/3\ \text{for}\ k=0,2,4$ for the MUB device and $\sum_k p_k=1$ for the SIC device).

\par
The comparison of QDT and QDSC in terms of the distribution of $\mathcal{L}$ is shown in Fig. \ref{fig:comparison}. The distribution reflects how well the range of the reconstruction fits the observed data, therefore give a DI verification of the reconstructions. It can be seen from the results that the QDSC shows less violations of Eq. (\ref{eq:inequality}) in average compared with QDT. In contrast with QDT which suffers from the errors in state preparation, the QDSC method is solely based on the measured statistics that is completely accessible at the detection side, thus is more robust to experimental imperfections in state preparation. The deviations from the bound in the results of QDSC are mainly attributed to the statistical fluctuations on the measurement results.
\par
The quantities $Q$ and $t$ represent the overlap between different elements of the POVM and the trace of each element respectively, which fully characterize the structure of the POVM, i.e. the physical model of the measurement device~\cite{PhysRevLett.123.140405}. The representation given by QDSC is up to symmetry transformations which can be understood as the transformation of the reference frame of the whole system~\cite{Wigner:102713,RevModPhys.79.555}. The reference frame can be specified with respect to the measurement POVM, then further usage of the measurement device can be conducted in a consistent way. To validate the usefulness of QDSC and demonstrate the break of the circular argument, we in turn perform a state tomography with the measurement calibrated by QDSC, and compare the results with those calibrated by QDT and those without priori calibration (see Fig. S2 of the Supplemental Material~\cite{Supplementary}).


\par
In conclusion, we realize quantum detector self-characterization, that solely utilizes the events produced in the measurement part to explore the geometrical structure of the detector response. We have applied the self-characterization method to two typical, extensively used measurements, highlighting its feasibility and robustness in practical cases. The present self-characterization method extends witness-based methods to a range-based method in characterizing quantum systems and devices. Together with a modelling on the response range of measurement operators, this method can be further generalized to more complicated devices. Future works will investigate the range for high-dimensional systems and entangled states. We expect the range-based techniques will become a new means for specifying quantum systems and mapping detector response~\cite{Cooper2014}, and find their applications in a wide range of quantum information tasks such as cryptography, random number generation~\cite{PhysRevApplied.7.054018} and metrology, especially where calibrating measuring apparatus is required in advance.


\begin{acknowledgments}
The authors thank M. Dall{$'$}Arno, F. Buscemi, B. Zeng, X. Ma, P. Zeng, Z. Hradil, L. L. S{\'a}nchez-Soto and N. Yu for enlightening discussions. This work was supported by the National Key Research and Development Program of China (Grant No. 2017YFA0303703) and National Natural Science Foundation of China (Grants No. 91836303, No. 61975077, No. 61490711, and No. 11690032).
\end{acknowledgments}

%
\clearpage
\begin{widetext}
\section{Supplemental Material}
\section{Experimental details}
\par
\textit{MUB and SIC measurements.--} Two orthogonal bases $\{|u_1\rangle,...,|u_d\rangle\}$ and $\{|v_1\rangle,...,|v_d\rangle\}$ are mutually unbiased for a $d$-dimensional quantum system if
$|\langle u_k|v_l\rangle|^2=1/d \quad \forall k,l$, while the SIC measurement~\cite{Renes2004} is described by the POVM elements $\pi_k=(1/d)\ket{\psi_k}\bra{\psi_k} \quad (k=0,1,2...d^2-1)$ satisfying $\mathrm{Tr}\left(\pi_{k} \pi_{l}\right)=1/[d^2(d+1)] \quad \text{for}\ k\neq l.$ For single qubit case $(d=2)$, the three bases of MUB measurement can be represented as $\{|0\rangle,|1\rangle\}$, $\{(|0\rangle+|1\rangle)/\sqrt{2},(|0\rangle-|1\rangle)/\sqrt{2}\}$, $\{(|0\rangle+i|1\rangle)/\sqrt{2},(|0\rangle-i|1\rangle)/\sqrt{2}\}$, where $|0\rangle,|1\rangle$ denote the horizontal and vertical polarization of the single photon respectively in the experiment. Correspondingly, the measurement operators of the SIC measurement can be written as
\begin{eqnarray}
\pi_0=\begin{pmatrix}
\frac{1}{2}  & 0 \\
0 & 0
\end{pmatrix},&&\
\pi_1=\begin{pmatrix}
\frac{1}{6}  & -\frac{\sqrt{2}}{3} \\
-\frac{\sqrt{2}}{3} & \frac{1}{3}
\end{pmatrix}, \nonumber \\
\pi_2=\begin{pmatrix}
\frac{1}{6}  & \frac{\sqrt{2}}{3}\mathrm{e}^{\mathrm{i}\pi/3} \\
\frac{\sqrt{2}}{3}\mathrm{e}^{-\mathrm{i}\pi/3} & \frac{1}{3}
\end{pmatrix},&&\
\pi_3=\begin{pmatrix}
\frac{1}{6}  & \frac{\sqrt{2}}{3}\mathrm{e}^{-\mathrm{i}\pi/3} \\
\frac{\sqrt{2}}{3}\mathrm{e}^{\mathrm{i}\pi/3} & \frac{1}{3}
\end{pmatrix}, \nonumber
\end{eqnarray}
in the $|0\rangle,|1\rangle$ basis.

\par
\textit{Experimental setup.--} Laser pulses ($\sim$150fs duration, 415nm central wavelength), frequency-doubled from a mode locked Ti:Sapphire laser (Coherent Mira-HP), pumped a phase-matched potassium di-hydrogen phosphate (KDP) crystal to generate photon pairs. The photon pairs were then separated into signal and idler modes by a polarizing beam-splitter (PBS). The detection of one photon in the idler mode heralds a single photon in the signal mode. The polarization of the single photons was manipulated by three waveplates with electronically-controlled rotation stages (Newport PR50PP) to generate sample states.
\par
For the measurement part, the transformations of the three bases of MUB are realized by a quater wave plate and a half wave plate, followed by a calcite beam-displacer (BD) and two single photon counting modules (Excelitas Technologies, SPCM-AQRH-FC). The SIC measurement is implemented by a photonic quantum walk network with BDs, wave plates and SPCMs~\cite{PhysRevLett.110.200404,PhysRevLett.114.203602,PhysRevA.91.042101}. The statistics of measurement outcomes were registered by a coincidence logic with a time window of 4.5ns. For each probe state, we collected data in about 1.4s and run the experiment 40 times to calculate the expectation value and standard uncertainty.
\par
\textit{Probe states.--} The 50 probe states used in the experiment are prepared by three electronically-controlled waveplates (quarter-half-quarter) following a polarizing beam splitter. The settings of the waveplates are configured to generate sample states $\{\rho^{(j)}\}$ of the form

\begin{gather}\label{Prob states}
  \frac{1}{2}(I+\sigma_z),\ \frac{1}{2}(I-\sigma_z),\nonumber \\
  \frac{1}{2}\left[I+\sin \left(\frac{l\pi}{4}\right) \cos \left(\frac{k\pi}{8}\right) \sigma_x+\sin \left(\frac{l\pi}{4}\right) \sin \left(\frac{k\pi}{8}\right)\sigma_y+\cos \left(\frac{l\pi}{4}\right)\sigma_z\right],\ \text{for }k=1,2,...,6;l=1,2,...,8.\nonumber
\end{gather}
In the experiment the actual prepared states may differ from the ideal states due to several forms of experimental imperfections. These systematic errors mainly stem from the misalignments of the optics axis for waveplates (typically $\sim$ 0.1 degree), the retardation errors of waveplates (typically $\sim \lambda /300$ where $\lambda=830\text{nm}$) and the inaccuracies of the rotation stages for waveplates (typically $\sim$ 0.025 degree). These factors may affect the tomographic results but not the results of the self-characterization method since the self-characterization does not rely on the exact form of probe states.

\section{Derivation of Equation (3)}
\par
Given a quantum measurement represented by a $n$-outcome POVM $\{\pi_k\}$, the constraints on the probability distribution $\bm{p}$ for an arbitrary quantum state stem from both the mapping $\{\pi_k\}$ and the set of the state space $\{\rho|\rho \geq 0,\ \Tr(\rho) = 1\}$. The allowed range of $\bm{p}$ is the set of expectation values of a set of Hermitian operators mapped from the state space,
$$\mathcal{W}(\pi):=\{(\Tr(\rho\pi_0),\Tr(\rho\pi_1),...,\Tr(\rho\pi_{n-1}))|\rho\geq 0, \Tr(\rho)=1\},$$ which is termed as joint algebraic numerical range (JANR) of Hermitian operators~\cite{Chen2015}. It has been shown that JANR is a convex and compact set and the JNRs of $2\times 2$ Hermitian observables are ellipses or (hyper-)ellipsoids \cite{Sehrawat2017}. For the single qubit case, a density operator $\rho$ and the elements $\{\pi_k\}$ of a POVM can be written in the Bloch representation as

\begin{equation}\label{rho}
  \rho=\frac{1}{2}(I+\mathbf{r}\cdot\bm{\sigma}),
\end{equation}
\begin{equation}\label{pi_y}
  \pi_k=t_kI+\bm{m}_k\cdot\bm{\sigma},
\end{equation}
respectively, where $\bm{\sigma}=(\sigma_x,\sigma_y,\sigma_z)$ is the tensor of Pauli operators and $\bm{r}=(r_x,r_y,r_z)$ is the Bloch vector of $\rho$. The operators $\pi_k$ can also be represented in a similar way by $t_k$ and $\bm{m}_k=(m_{k,x},m_{k,y},m_{k,z})$. Here a physical state should satisfy the positivity constraint $|\bm{r}|^2\leq1$. On the other hand, the requirement of a POVM $\sum\pi_k=I$ implies that $ m_{k,x}+m_{k,y}+m_{k,z}=0$ and $\sum_k t_k=1$, while the positivity constraint $\pi_k\geq 0$ implies $t_k>|\bm{m}_k|$. Given such a representation of $\rho$ and $\{\pi_k\}$, according to Born's rule we have
\begin{equation}\label{Born_rule}
  p_k=\Tr(\rho\pi_k) = t_k+ \bm{m}_k \cdot \bm{r}.
\end{equation}
Equation (\ref{Born_rule}) can be written in a matrix form
\begin{gather}\label{matrix_form}
  \begin{pmatrix} p_0-t_0 \\ p_1-t_1\\ ... \\ p_{n-1}-t_{n-1} \end{pmatrix} =
  \begin{pmatrix} m_{0,x} & m_{0,y} & m_{0,z} \\ m_{1,x} & m_{1,y} & m_{1,z} \\ ... \\ m_{n-1,x} & m_{n-1,y} & m_{n-1,z} \end{pmatrix}\cdot
  \begin{pmatrix} r_x \\ r_y \\ r_z  \end{pmatrix},
\end{gather}
or denoted as $(\bm p-\bm t)=M\cdot \bm r$, where $M$ is an $n\times3$ matrix. In this form, the positivity constraint $|\bm{r}|^2\leq 1$ of $\rho$ can be recast into a constraint on $\bm p$ since $|\bm{r}|^2=\bm{r}^T \bm{r} =(\bm p-\bm t)^T(M^+)^TM^+(\bm p-\bm t)$, where $M^+$ is the Moore-Penrose pseudoinverse of $M$ satisfying $MM^+M=M$. By defining $Q=MM^T$ and thereby $Q^+=(MM^T)^+=(M^T)^+M^+$, we arrive at
 \begin{equation}\label{centre_form_of_p}
   (\bm p-\bm t)^TQ^+(\bm p-\bm t)\leq1,
 \end{equation}
which is in a centre form of an $n$ dimensional hyper-ellipsoid. This indicates that $\bm p$ should lie in the hyper-ellipsoid determined by the matrix $Q^+$ centered at $\bm t$. $Q$ and $\bm t$ can be given by the POVM as
\begin{equation}\label{Qij}
  \begin{cases}
  Q_{k,l}=\bm{m}_k^T\bm{m}_l=\frac{1}{2}\Tr (\pi_k\pi_l)-\frac{1}{4}\Tr (\pi_k)\Tr (\pi_l),\\
  t_k=\frac{1}{2}\Tr(\pi_k).
  \end{cases}
\end{equation}
Since four linear independent $2\times 2$ matrices form a complete basis, and a POVM requires the normalization constraint $\sum_k t_k=1$, the number of linear independent elements of a qubit POVM can not be larger than 3. Therefore the set $\mathcal{W}(\pi)$ should be an ellipsoid lying on a 3-dimensional affine plane. The equality sign in Eq. (3) holds only when $|\bm{r}|^2=\Tr(\rho^2)=1$, which states that extreme points of the JANR are obtained by pure states. For any orthogonal transformation $O$ satisfying $O^TO=\openone$ and the map is trace-preserving, $\{O\pi_k O^T\}$ gives the equivalent $Q$ and $\bm t$. In this sense, $Q$ and $\bm t$ determine a equivalence class of POVMs up to unitary and anti-unitary operations. 
\par
Given $Q$ quantifies the overlap between different elements of the POVM and $\bm t$ quantifies the trace of each element, $Q$ and $\bm t$ is a representation of the structure of the POVM elements which completely characterizes the physical model of the measurement device, including the experimental errors in the measurement operators. According to Wigner's theorem~\cite{Wigner:102713}, for a transformation $O$ on the Hilbert space if $|\langle\phi|O^T O|\psi\rangle|=|\langle\phi |\psi\rangle|$ for any $|\psi\rangle$ and $|\phi\rangle$, then $O$ is an unitary or anti-unitary transformation. Such a symmetry transformation can be understood as a rotation of our view on the whole system, i.e. a reference frame that the description of the system is relative to~\cite{RevModPhys.79.555}. In other words, the physical model remains the same under symmetry transformations which do not change the measurement outcomes~\cite{PhysRevLett.123.140405}. In our case to characterize a detector for further usage, the reference frame can be defined by the side of the measurement in the characterization (for example the coordinate in the reconstructed range in Figs. 4(b) and (d) left can be naturally regarded as a reference frame of the measurement), then further usage can be conducted in terms of this reference frame.
\par

\section{Characterization procedure}
\par
For an $n$-outcome qubit POVM probed with $m$ states, we obtain a data set of the measured statistics $\{\bm{p}^{(j)}\}$ which can be described by an $n \times m$ matrix $P$
\begin{equation}
P_{n \times m}=
\begin{pmatrix}
p_0^{(0)} & p_0^{(1)} &\cdots & p_0^{(m-1)} \\
p_1^{(0)} & p_1^{(1)} &\cdots & p_1^{(m-1)} \\
\vdots& \vdots& \ddots &\vdots  \\
p_{n-1}^{(0)} &p_{(n-1)}^{(1)} &\cdots & p_{n-1}^{(m-1)}
\end{pmatrix}. \nonumber
\end{equation}

Here the $m$ columns of $P$ represent the statistics of different probe states and each column corresponds to an observed probability distribution $\bm{p}^{(j)}$. The QDSC procedure consists of three steps:
\par
(i) Extract features in the data. The aim of this step is to remove the redundant linear dependencies among the measurement outcomes in the presence of statistical fluctuations. First, subtract the average probability over the $m$ probe states $\bar{\bm{p}}=(\bar{p}_0,\bar{p}_1,...,\bar{p}_{n-1})$ from the matrix $P$, where 
$$\bar{p}_k=\frac{1}{m}\sum_{j=0}^{m-1} p_k^{(j)}.$$ 
As a result, we obtain a new data set $A_{n \times m}$
\begin{equation}
A_{n \times m}=
\begin{pmatrix}
p_0^{(0)}-\bar{p}_0 & p_0^{(1)}-\bar{p}_0 &\cdots & p_0^{(m-1)}-\bar{p}_0 \\
p_1^{(0)}-\bar{p}_1 & p_1^{(1)}-\bar{p}_1 &\cdots & p_1^{(m-1)}-\bar{p}_1 \\
\vdots& \vdots& \ddots &\vdots  \\
p_{n-1}^{(0)}-\bar{p}_{n-1} &p_{(n-1)}^{(1)}-\bar{p}_{n-1} &\cdots & p_{n-1}^{(m-1)}-\bar{p}_{n-1}
\end{pmatrix}. \nonumber
\end{equation}
This procedure introduces the normalization condition $\sum_k p_k^{(j)}=1$ into data processing without changing the linear dependencies among measurement outcomes, due to the fact the average probability $\bar{\bm{p}}$ has the same linear dependence as the raw data set $P$. We have $\text{rank}(A)=\text{rank}(P)-1$. Then, perform singular value decomposition on the matrix $A=U\cdot \Sigma\cdot V^T$, which is in general of the form
\begin{equation}
A_{n \times m}=U_{n \times n}
\begin{pmatrix}
s_1 & & & \quad & \\
& \ddots & &\quad &  \\
& & s_n &\quad &
\end{pmatrix}_{n \times m}
  V^T_{m \times m}. \nonumber
\end{equation}
Due to the linear dependencies in the measurement operators, the response range should lie in an affine plane up to 3 dimensional
(3 dimensional for informationally complete measurements, see the last section for an example of incomplete measurement). In general cases this leads to up to 3 large singular values, and other singular values should be zero or of the order $O(1/\sqrt{N})\ll s_1(s_2,s_3)$, here $N$ is the sum of the collected counts of measurement outcomes $N=\sum_k n_k$ for each state. Thus we have
\begin{equation}
A_{n \times m}\approx U_{n \times 3}
\begin{pmatrix}
s_1 & &  \\
& s_2 &   \\
& & s_3
\end{pmatrix}
  V^T_{3 \times m}. \nonumber
\end{equation}
The reduced data $\tilde{A}$ can be obtained by
\begin{equation}
\tilde{A}_{3 \times m}={(U^T)}_{3 \times n}\cdot A_{n \times m}\approx
\begin{pmatrix}
s_1 & &  \\
& s_2 &   \\
& & s_3
\end{pmatrix}
  V^T_{3 \times m}. \nonumber
\end{equation}
This is virtually a principle component analysis of the data set robust to experimental noise and we use the reduced data $\tilde{A}$ to conduct the following procedure.
\par
(ii) Get the boundary data set $B$ lying on the boundary of the convex hull of the reduced data $\tilde{A}$, that is,
\begin{equation}
B=\{\bm{v}_j=(\tilde{A}(0,j),\tilde{A}(1,j),...,\tilde{A}(n-1,j))^T,\text{ for }\bm{v}_j\in\text{the boundary of ConvexHull}(\{\bm{v}_j|j=0,1,...,m-1\})\}. \nonumber
\end{equation}
Here $\tilde{A}(i,j)$ stands for the element in the $(i+1)$-th row, $(j+1)$-th column of the matrix $\tilde{A}$. This boundary data set can be described by a matrix $B_{n \times m'}$, where $m'$ is the number of boundary points on the convex hull of $\tilde{A}$. The boundary data can be calculated by a standard algorithm such as the Delaunay Triangulation method. Given that the response range $\mathcal{W}(\pi)$ is a convex set and the interior points of the set can be easily obtained by linear combinations of boundary points, we need to get the boundary data in the whole data set if we are to reconstruct the geometrical shape of the convex set.  
\par
(iii) Reconstruct an ellipsoid in the space $\tilde{p}=U\cdot (p-\bar{p})$ via a direct ellipsoid fitting fed with $\tilde{A'}$ under the physical constraint. This is equivalent to solving the optimization problem in Eq. (4) in the main text since $\tilde{p}$ and $p$ are related via only a linear map $U$. Finally, map the estimated results to the whole response range $\mathcal{W}(\pi)$ and output the corresponding $Q$ and $t$.
\par
The optimization problem in Eq.~(4) is in general not a convex optimization. In the optimization we adopt a global search algorithm to overcome the non-convexity of the problem. The algorithm runs constrained nonlinear optimization local solvers with iteratively-generated initial points to find the global minimum. The algorithm outputs an optimization result when the local solvers converged to a global minimum. The algorithm works well in our case and the results all show high fidelities with the results given by QDT. We anticipate further developments on the algorithm aspect to solve this optimization problem.
\par
Another question in the characterization procedure is how many probe states are required to give a reliable estimation $\{Q,\bm t\}$. In the present case where the QDSC procedure is to estimate an ellipsoid with 9 unknown parameters, the solution can be determined by a system of 9 linear equations (under certain conditions). Therefore 9 probe states are necessary to have a reliable reconstruction of the response range. This required number of states is more than that for QDT, i.e. the tomographically complete set of probe states (4 states for a qubit system) due to the QDSC procedure relaxes priori knowledge on probe states. See Fig.~\ref{fig:reduce} in the following section for a detailed analysis on the performance of QDSC with different number of probe states.
\section{Detailed results}

\par
The detailed experimental results of both QDSC and QDT for our two measurement devices are
\begin{equation}
Q^{\rm{sc}}=\begin{pmatrix}
0.0276 & -0.0276 & 0.0005 & -0.0005 & -0.0001 & 0.0001\\
-0.0276 & 0.0276 & -0.0005 & 0.0005 & 0.0001 & -0.0001\\
0.0005 & -0.0005 & 0.0276 & -0.0276 & -0.0001 & 0.0001\\
-0.0005 & 0.0005 & -0.0276 & 0.0276 & 0.0001 & -0.0001 \\
-0.0001 & 0.0001 & -0.0001 & 0.0001 & 0.0277 & -0.0277 \\
0.0001 & -0.0001 & 0.0001 & -0.0001 & -0.0277 & 0.0277
\end{pmatrix},\quad
t^{\rm{sc}}=\begin{pmatrix}
0.1661 \\ 0.1672 \\ 0.1661 \\ 0.1673 \\0.1663 \\ 0.1670
\end{pmatrix}, \nonumber
\end{equation}

\begin{equation}
Q^{\rm{tomo}}=\begin{pmatrix}
0.0273 & -0.0273 & 0.0003 & -0.0003 & -0.0001 & 0.0001\\
-0.0273 & 0.0273 & -0.0003 & 0.0003 & 0.0001 & -0.0001\\
0.0003 & -0.0003 & 0.0273 & -0.0273 & -0.0000 & 0.0000\\
-0.0003 & 0.0003 & -0.0273 & 0.0273 & 0.0000 & -0.0000 \\
-0.0001 & 0.0001 & -0.0000 & 0.0000 & 0.0274 & -0.0274 \\
0.0001 & -0.0001 & 0.0000 & -0.0000 & -0.0274 & 0.0274
\end{pmatrix},\quad
t^{\rm{tomo}}=\begin{pmatrix}
0.1654 \\ 0.1679 \\ 0.1653 \\ 0.1680 \\0.1657 \\ 0.1677
\end{pmatrix}, \nonumber
\end{equation}

for the MUB device, and
\begin{equation}
Q^{\rm{sc}}=\begin{pmatrix}
0.0616  & -0.0211 & -0.0182 & -0.0223 \\
-0.0211 & 0.0584  & -0.0217 & -0.0155 \\
-0.0182 & -0.0217 & 0.0611  & -0.0212 \\
-0.0223 & -0.0155 & -0.0212 & 0.0591
\end{pmatrix},\quad
t^{\rm{sc}}=\begin{pmatrix}
0.2494 \\ 0.2454 \\ 0.2535 \\ 0.2517
\end{pmatrix}, \nonumber
\end{equation}

\begin{equation}
Q^{\rm{tomo}}=\begin{pmatrix}
0.0618  & -0.0204 & -0.0184 & -0.0230 \\
-0.0204 & 0.0578  & -0.0223 & -0.0151 \\
-0.0184 & -0.0223 & 0.0614  & -0.0207 \\
-0.0230 & -0.0151 & -0.0207 & 0.0589
\end{pmatrix},\quad
t^{\rm{tomo}}=\begin{pmatrix}
0.2487 \\ 0.2423 \\ 0.2538 \\ 0.2552
\end{pmatrix}, \nonumber
\end{equation}
for the SIC device. The corresponding infidelities for the results of QDT and QDSC are shown in Table \ref{tab:fidility}. Here the definition of fidelity for $Q$ and $t$ is given by $F(Q^{\text{tomo}},Q^{\text{sc}})=\left[\Tr\left(\sqrt{Q^{\text{tomo}}}Q^{\text{sc}}\sqrt{Q^{\text{tomo}}}\right)\right]^2/\left[\Tr\left(Q^{\text{tomo}}\right)\Tr\left(Q^{\text{sc}}\right)\right]$ and $F(t^{\text{tomo}},t^{\text{sc}})=\left(\sum_k \sqrt{t_k^\text{tomo}t_k^\text{sc}}\right)^2$. 
\begin{table}[h!]
 \centering \caption{\label{tab:fidility}Infidelities $\log_{10}(1-F)$ between QDT and QDSC.}
\begin{tabular}{ccc}
\hline
    &\textbf{MUB device} & \textbf{SIC device}\\
    \hline
    $\log_{10}[1-F(Q^{\text{tomo}},Q^{\text{sc}})]$ & $-4.9330$ & $-4.2765$\\
    $\log_{10}[1-F(t^{\text{tomo}},t^{\text{sc}})]$ & $-5.3344$ & $-4.6549$\\
    \hline
   \end{tabular}
    \end{table}
\par
As we noted in the main text and the previous section, we can give a representation of the measurement POVM by specifying a reference frame. This can also be done by an ancillary experimental procedure to align the reference frame. As an example, we choose a particular reference frame to match the basis of the polarization of the photon in the experimental realization of the measurement devices for the following comparison with the results given by QDT. Back to the Bloch sphere representation of quantum states and measurement operators in Eqs.~(\ref{rho}) and (\ref{pi_y}), for the SIC device in our experiment we specify a reference frame that the $z$ direction of the Bloch sphere to be parallel with the vector $\bm{m}_0$, and then set the $x-z$ plane of the Bloch sphere to be the plane determined by the two vectors $\bm{m}_0$ and $\bm{m}_1$. This is equivalent to let $m_{0,x}=m_{0,y}=0$ and $m_{1,y}=0$ in Eq.~(\ref{pi_y}). As a result, the basis diagonalizes the operator for the first outcome $\pi_0^{\text{sc}}$, and the operator for the second outcome $\pi_1^{\text{sc}}$ is a real operator. Corresponding to the photonic system to realize the measurement devices, such an alignment equivalently determines the horizontal and vertical polarization of the photon by the measurement outcomes. Then other operators of the POVM are determined as $Q^{\text{sc}}$ and $\bm t^{\text{sc}}$ have well characterized the overlap between different measurement operators and the trace of each operators. By deriving the values of other elements in the matrix $M$ for the SIC device with reconstructed $Q^{\text{sc}}$, we get the following representation of $\{\pi_k^{\text{sc}}\}$
\begin{eqnarray}
\pi_0^{\text{sc}}=\begin{pmatrix}
0.4977  & 0 \\
0 & 0.0011
\end{pmatrix},&&\
\pi_1^{\text{sc}}=\begin{pmatrix}
0.1602  & -0.2261 \\
-0.2261 & 0.3305
\end{pmatrix}, \nonumber \\
\pi_2^{\text{sc}}=\begin{pmatrix}
0.1803  & 0.1237+0.2012\mathrm{i} \\
0.1237-0.2012\mathrm{i} & 0.3267
\end{pmatrix},&&\
\pi_3^{\text{sc}}=\begin{pmatrix}
0.1617  & 0.1024-0.2012\mathrm{i} \\
0.1024+0.2012\mathrm{i} & 0.3417
\end{pmatrix}. \nonumber
\end{eqnarray}
Similarly, for the MUB device, by setting the $z$ direction of the Bloch sphere to be parallel with the vector $\bm{m}_4$ and the $x-z$ plane of the Bloch sphere to be the plane determined by the two vectors $\bm{m}_0$ and $\bm{m}_4$, we derive the following representation of the MUB device:
\begin{eqnarray}
\pi_0^{\text{sc}}=\begin{pmatrix}
0.1654  & 0.1661 \\
0.1661 & 0.1669
\end{pmatrix},&&\
\pi_1^{\text{sc}}=\begin{pmatrix}
0.1679  & -0.1661 \\
-0.1661 & 0.1665
\end{pmatrix}, \nonumber \\
\pi_2^{\text{sc}}=\begin{pmatrix}
0.1656  & 0.0031-0.1660\mathrm{i} \\
0.0031+0.1660\mathrm{i} & 0.1665
\end{pmatrix},&&\
\pi_3^{\text{sc}}=\begin{pmatrix}
0.1677  & -0.0031+0.1660\mathrm{i} \\
-0.0031-0.1660\mathrm{i} & 0.1668
\end{pmatrix}. \nonumber  \\
\pi_4^{\text{sc}}=\begin{pmatrix}
0.3326  & 0 \\
0 & 0
\end{pmatrix},&&\
\pi_5^{\text{sc}}=\begin{pmatrix}
0.0008  & 0 \\
0 & 0.3333
\end{pmatrix}. \nonumber
\end{eqnarray}
These results are close to the corresponding tomographic reconstructions:
\begin{eqnarray}
\pi_0^{\text{tomo}}=\begin{pmatrix}
0.4971  & 0.0105+0.0028\mathrm{i} \\
0.0105-0.0028\mathrm{i} & 0.0002
\end{pmatrix},&&\
\pi_1^{\text{tomo}}=\begin{pmatrix}
0.1509  & -0.2222-0.0069\mathrm{i} \\
-0.2222+0.0069\mathrm{i} & 0.3338
\end{pmatrix}, \nonumber \\
\pi_2^{\text{tomo}}=\begin{pmatrix}
0.1827  & 0.1231+0.2029\mathrm{i} \\
0.1231-0.2029\mathrm{i} & 0.3249
\end{pmatrix},&&\
\pi_3^{\text{tomo}}=\begin{pmatrix}
0.1694  & 0.1096-0.1988\mathrm{i} \\
0.1096+0.1988\mathrm{i} & 0.3411
\end{pmatrix}, \nonumber
\end{eqnarray}
for the SIC device, and
\begin{eqnarray}
\pi_0^{\text{tomo}}=\begin{pmatrix}
0.1640  & 0.1652-0.0078\mathrm{i} \\
0.1652+0.0078\mathrm{i} & 0.1668
\end{pmatrix},&&\
\pi_1^{\text{tomo}}=\begin{pmatrix}
0.1693  & -0.1652+0.0078\mathrm{i} \\
-0.1652-0.0078\mathrm{i} & 0.1666
\end{pmatrix}, \nonumber \\
\pi_2^{\text{tomo}}=\begin{pmatrix}
0.1653  & -0.0059-0.1652\mathrm{i} \\
-0.0059+0.1652\mathrm{i} & 0.1653
\end{pmatrix},&&\
\pi_3^{\text{tomo}}=\begin{pmatrix}
0.1680  & 0.0059+0.1652\mathrm{i} \\
0.0059-0.1652\mathrm{i} & 0.1680
\end{pmatrix}. \nonumber  \\
\pi_4^{\text{tomo}}=\begin{pmatrix}
0.3313  & 0.0009 \\
0.0009 & 0
\end{pmatrix},&&\
\pi_5^{\text{tomo}}=\begin{pmatrix}
0.0020  & -0.0009 \\
-0.0009 & 0.3333
\end{pmatrix}, \nonumber
\end{eqnarray}
for the MUB device. The fidelities between POVM elements given by QDSC and QDT are shown in Table \ref{tab:povm_Fidelity}, where $F(\pi_k^{\text{tomo}},\pi_k^{\text{sc}})$ is the fidelity between different POVM elements defined as 
$$F(\pi_k^{\text{tomo}},\pi_k^{\text{sc}})=\left[\Tr(\sqrt{\pi_k^{\text{tomo}}}\pi_k^{\text{sc}}\sqrt{\pi_k^{\text{tomo}}})\right]^2/\left[\Tr(\pi_k^{\text{tomo}})\Tr(\pi_k^{\text{sc}})\right].$$ 
The result indicates the POVMs estimated by the QDSC and QDT are very close to each other even with the presence of a potential mismatch of the reference frames in the two representations. Note such a mismatch of the reference frame between different characterization methods can also be eliminated by applying the same procedure of specifying the reference frame to the different reconstructions.
\begin{table}[h!]
  \centering    \caption{\label{tab:povm_Fidelity}Fidelities of the POVM elements between QDSC and QDT for the MUB device and SIC device.}
\begin{tabular}{ccccccc}
      \hline
      $F(\pi_k^{\text{tomo}},\pi_k^{\text{sc}})$ & $\pi_0$ & $\pi_1$ & $\pi_2$ & $\pi_3$ & $\pi_4$ & $\pi_5$ \\
      \hline
      \textbf{MUB device} & 99.94 & 99.85 & 99.92 & 99.83 & 100.00 & 99.91 \\
      \textbf{SIC device} & 99.73 & 99.90 & 100.00 & 99.90 &   &   \\
      \hline
      
 \end{tabular}
\end{table}
\begin{figure}
\centering
\addtocounter{figure}{1}
\subfigure{\includegraphics[width=0.45\linewidth]{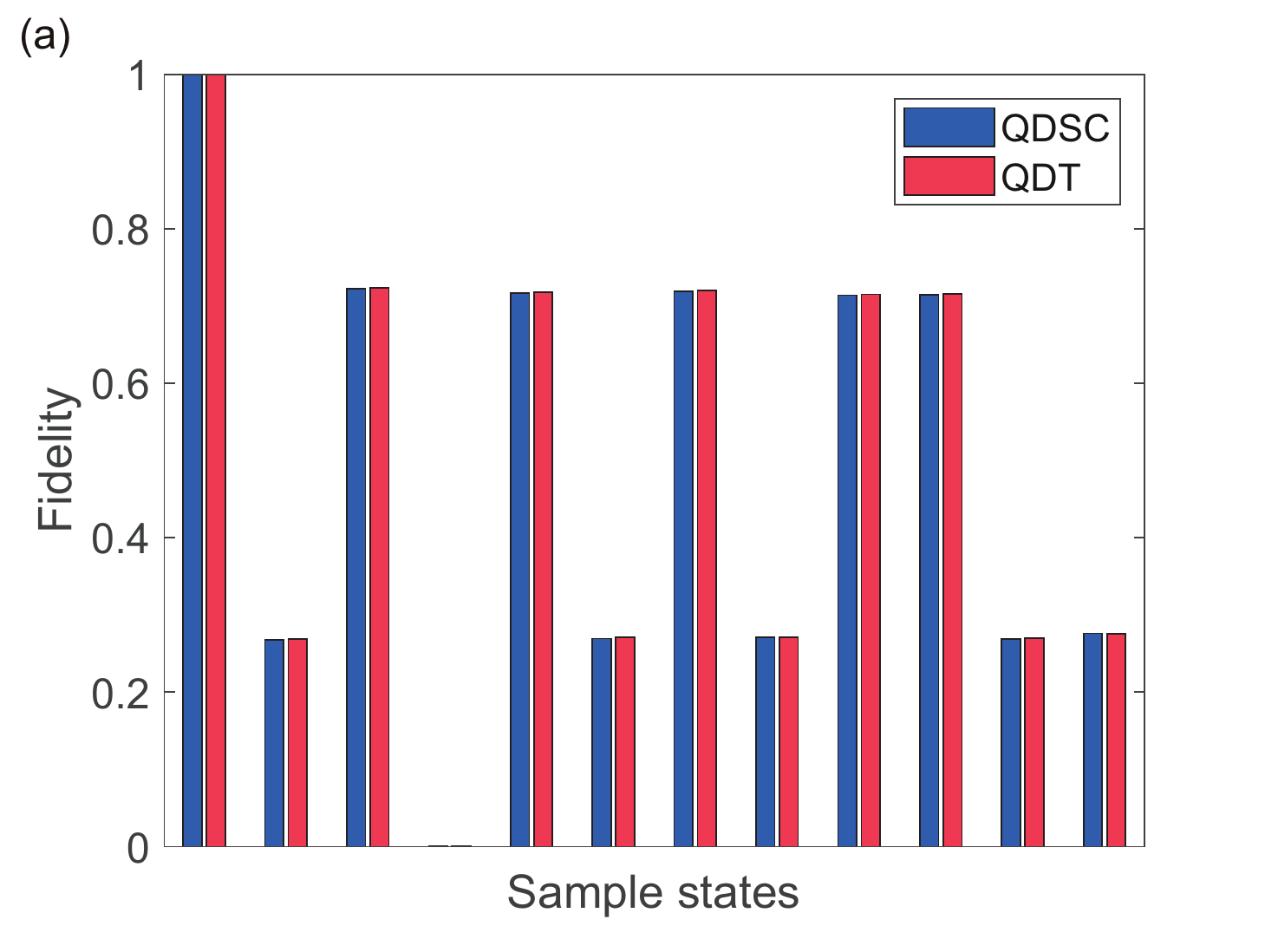}\label{fig:overlap:a}}
\quad
\subfigure{\includegraphics[width=0.45\linewidth]{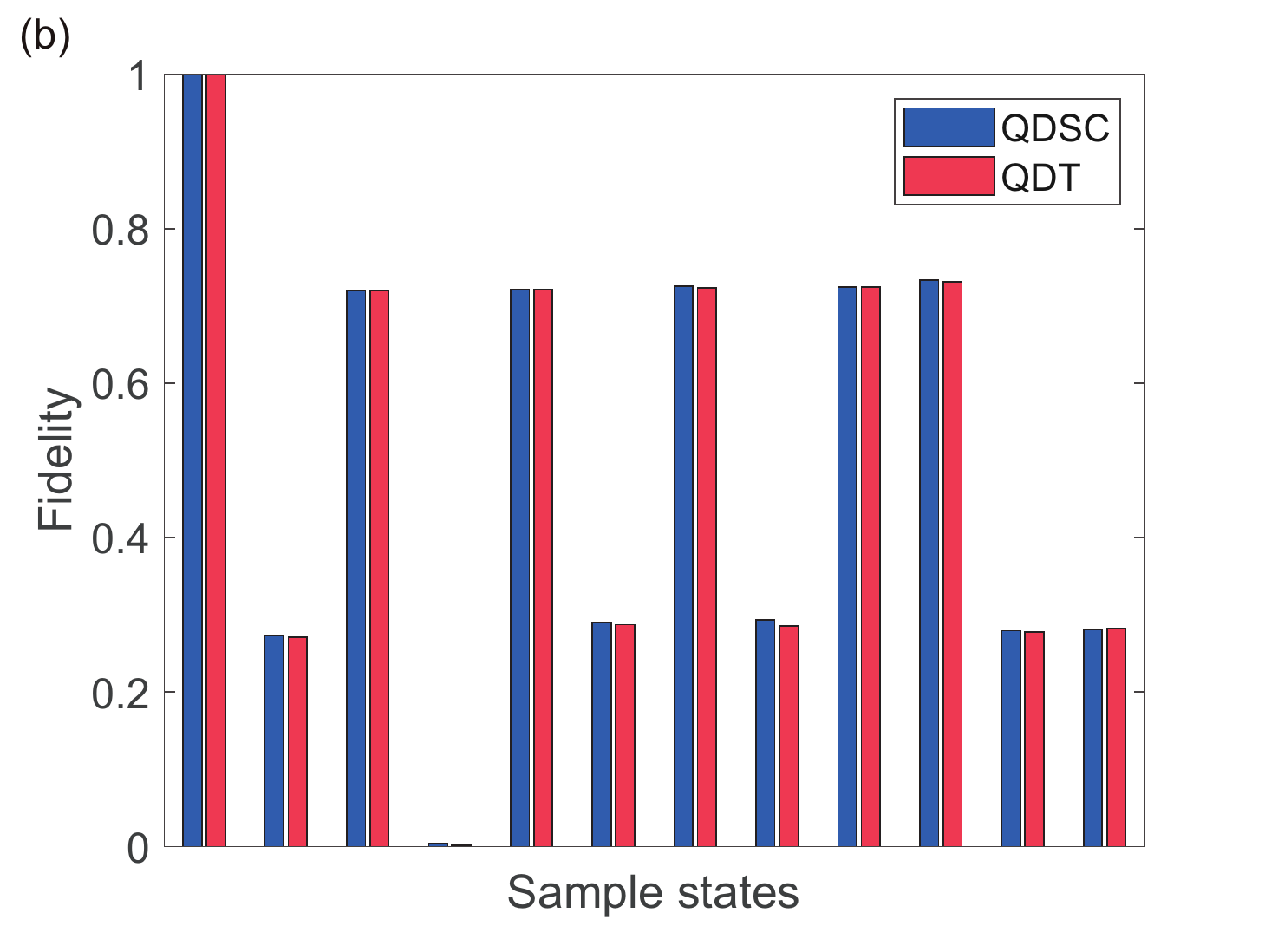}\label{fig:overlap:b}}
\addtocounter{figure}{-1}
\caption{\label{fig:overlap} The average fidelities of the reconstructions of the 12 sample states with the first state, which quantify the overlap between different states. The reconstructions are obtained by state tomography performed with the measurement calibrated by different methods: QDSC (blue) and QDT (red). (a) The results for the MUB device; (b) the results for the SIC device.}
\end{figure}

\begin{figure}
\centering
\addtocounter{figure}{1}
\subfigure{\includegraphics[width=0.45\linewidth]{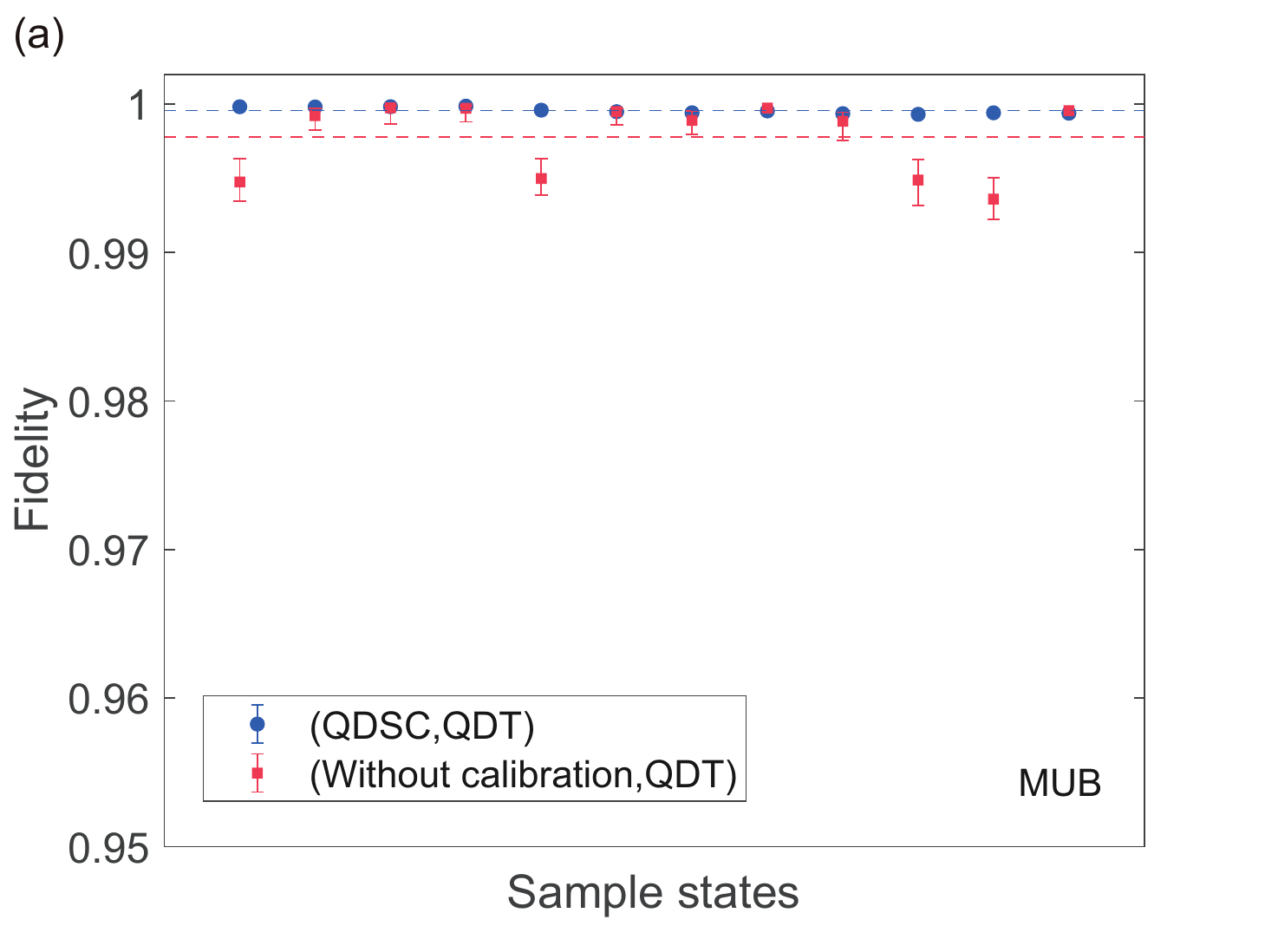}\label{fig:qst:a}}
\quad
\subfigure{\includegraphics[width=0.45\linewidth]{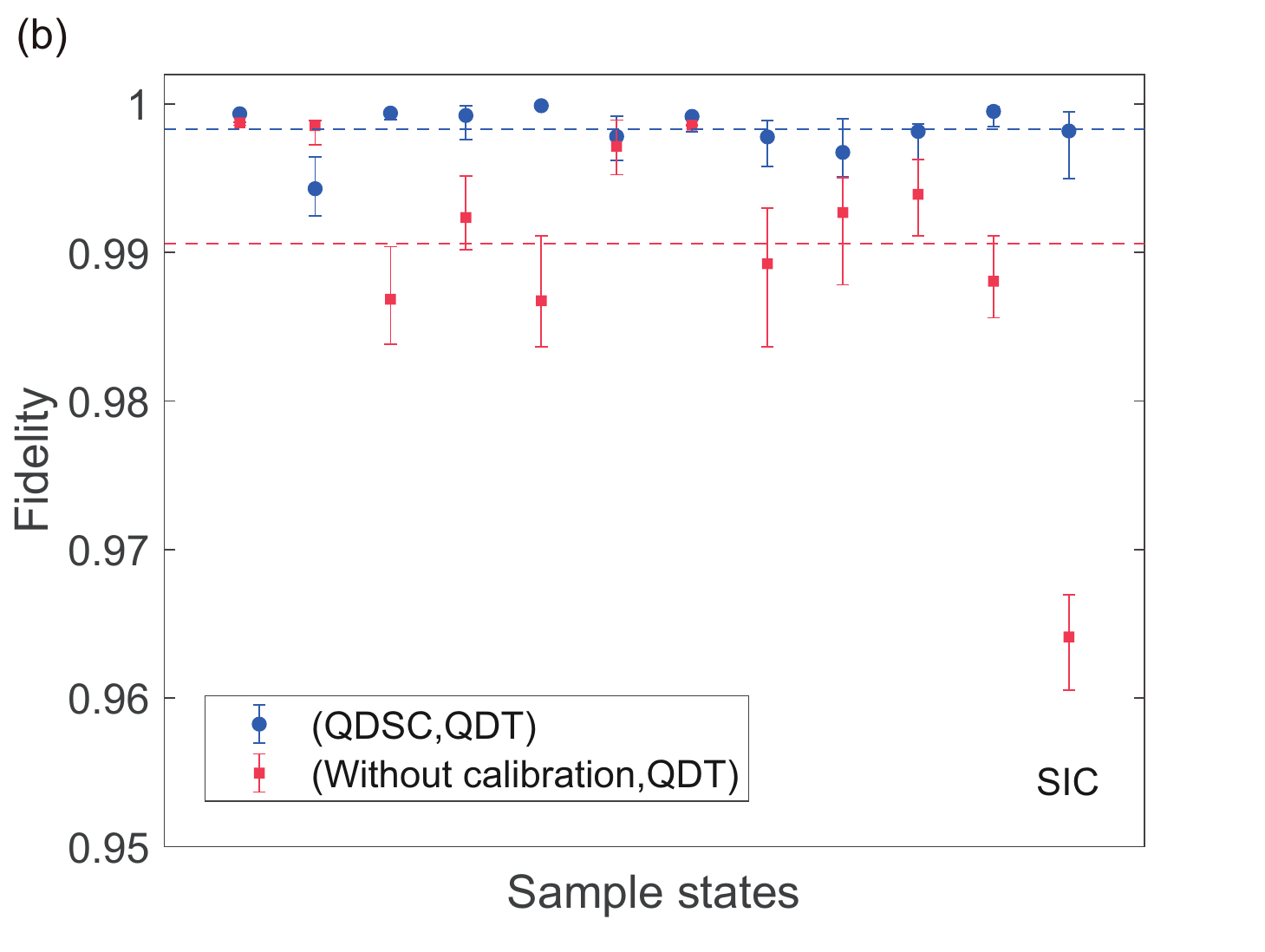}\label{fig:qst:b}}
\addtocounter{figure}{-1}
\caption{\label{fig:qst} The reconstruction fidelities of state tomography for 12 sample states, performed with the measurement calibrated by different methods: QDSC compared with QDT (blue dots) and the measurement without priori calibration compared with QDT (red squares). (a) The results for the MUB device; (b) the results for the SIC device. The dashed lines represent the fidelity averaged over the 12 states. The average fidelities of state tomography evidently increase after a QDSC compared with those without a priori calibration. The errorbars are the standard deviations of the fidelities over 40 runs of the state tomography experiment.}
\end{figure}
\par
To demonstrate the break of the circular argument in characterizing quantum systems and devices, we reversely perform a quantum state tomography with the measurement scheme calibrated by QDSC $\{\pi_k^{\text{sc}}\}$. The tomography is conducted with 12 uniformly-sampled states that form an icosahedron in Bloch sphere (different from the 50 states used to characterize the measurement device). The results are compared with those with the measurement scheme calibrated by QDT $\{\pi_k^{\text{tomo}}\}$ and analyzed in two ways. The first one is a basis-independent way to look at the fidelities between different states, which are shown in Fig.~\ref{fig:overlap}. The fidelity between two states are defined as $$F(\rho^{(j)},\rho^{(k)})=\left[\Tr(\sqrt{\rho^{(j)}}\rho^{(k)}\sqrt{\rho^{(j)}})\right]^2.$$
We plot the fidelities between one of the 12 states and the first state $F(\rho^{(j},\rho^{(0)})$ which quantify the relative overlap between these states. The fidelities are in well agreements with those performed by QDT. The second one is to analyze the fidelities of the reconstructions for the same state but given by different methods, as shown in Fig.~\ref{fig:qst}. We plot the fidelities of states, in comparison with the results where the measurement scheme is calibrated by QDT, before and after the QDSC procedure. The results show that the reconstruction fidelities increase evidently after a self-characterization, compared with those without priori calibration and resort to an ideal POVM as described in the part ``\textit{MUB and SIC measurements}'' of the first section. This result validates the performance of the QDSC method in a usage of the calibrated measurement.

\begin{figure}
\centering
\addtocounter{figure}{1}
\subfigure{\includegraphics[width=0.45\linewidth]{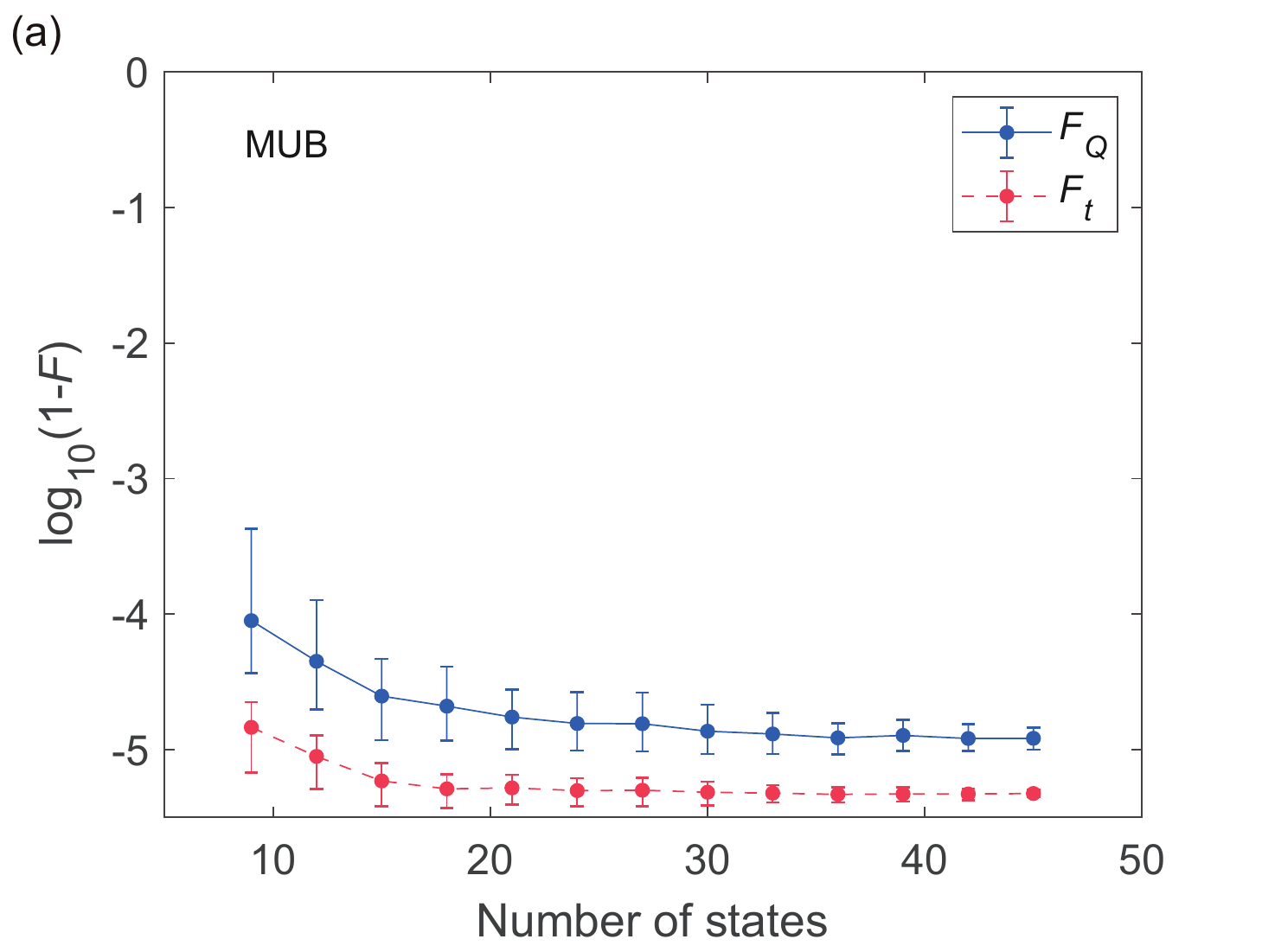}\label{fig:reduce:a}}
\quad
\subfigure{\includegraphics[width=0.45\linewidth]{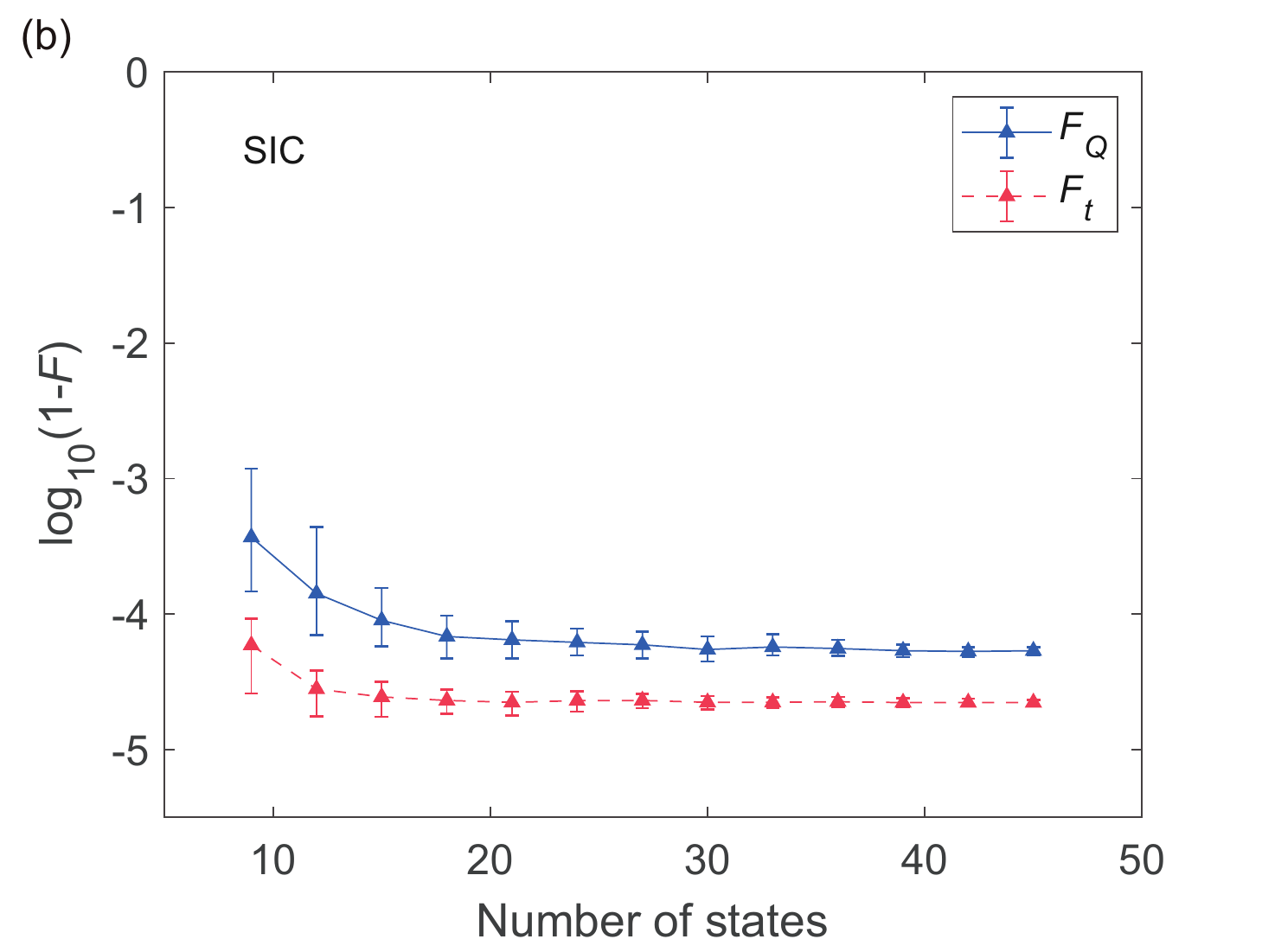}\label{fig:reduce:b}}
\addtocounter{figure}{-1}
\caption{\label{fig:reduce} Average infidelities vs. number of sample states for QDSC with randomly sampled states. (a) The results for the MUB device; (b) the results for the SIC device. The errorbars are the standard deviations of the fidelities over 100 runs of the sample procedure.}
\end{figure}
\par
To further investigate the performance and the scale of the self-characterization method, we conduct the characterization with reduced probe states randomly chosen from the overall 50 state preparations. We randomly generate $m$ states from the 50 probe states to perform the characterization and repeat the procedure $100$ times to obtain the average fidelities. And we change the number of states $m$ from 9 to 45 to show the performance of the method. Figure \ref{fig:reduce} shows the results for the MUB device and SIC device. It can be concluded that the average infidelities decrease with the increase of the number of states and converge to the level of the results with 50 states.

\section{Incomplete measurements}
\begin{figure}
    \includegraphics[width=\textwidth]{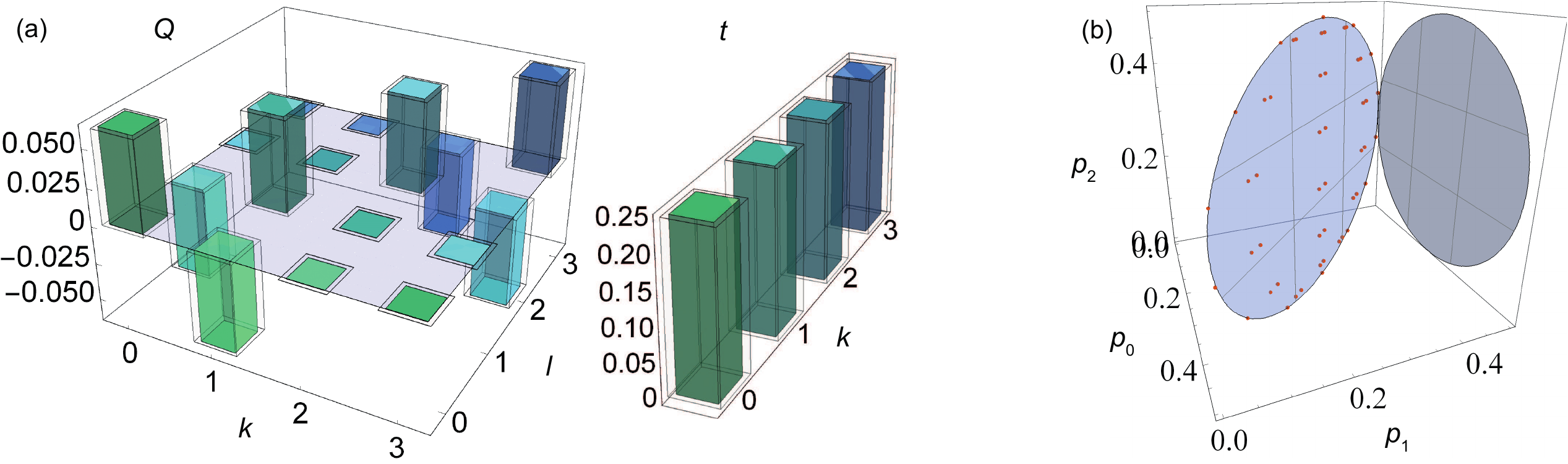}
  \caption{\label{fig:DIreal}Results of quantum detector self-characterization (QDSC) for the incomplete measurement, real MUB measurement. (a) The reconstructed $Q$ and $t$ (chromatic bars). The corresponding results of quantum detector tomography (transparent bars with solid line edge) are also plotted for comparison. (b) The estimated response range (blue region) and the measured data, illustrated in the probability space.} 
\end{figure}
\par
The MUB and SIC measurements used in the main text are both informationally-complete measurements, that is, every state can be completely determined by the statistics of the measurement. Yet in a more general scenario, there are also incomplete measurements by which it is insufficient to completely infer the state. The self-characterization of incomplete measurements is the same as that for the complete ones, as long as the probe states can sufficiently recover the range of the measurement. In the following we show the self-characterization of an incomplete measurement, real MUB measurement for qubit system, which contains only 2 bases constrained in real projectors. The results are shown in Fig. \ref{fig:DIreal}. The self-characterization method still work well for this incomplete measurement. The difference is that for this incomplete measurement, the linear independent dimension is reduced to 2, shown as an ellipse instead of an ellipsoid in the probability space.
\end{widetext}
\end{document}